\documentclass[final]{aipproc}
\layoutstyle{6x9}

\newcommand{\beq}{\begin{equation}}
\newcommand{\eeq}{\end{equation}}
\newcommand{\beqa}{\begin{eqnarray}}
\newcommand{\eeqa}{\end{eqnarray}}
\newcommand{\bea}{\begin{array}}
\newcommand{\eea}{\end{array}}
\newcommand{\Pexp}{{\rm Pexp}}

\newcommand{\Tr}{{\rm Tr}}

\newcommand{\diag}{{\rm diag}}

\newcommand{\re}{\relax{\rm I\kern-.18em R}}

\newcommand{\pl}{{{\cal P}_\infty}}

\begin{document}

\title{A Review of Instanton Quarks and Confinement}

\classification{11.10.Wx, 12.38.Aw, 14.80.Hv}
\keywords{Calorons, Confinement, QCD}

\author{Pierre van Baal}{
  address={Instituut-Lorentz for Theoretical Physics, University of Leiden,\\
           P.O.Box 9506, NL-2300 RA Leiden, The Netherlands}
}

\begin{abstract}
We review the recent progress made in understanding
instantons at finite temperature (calorons) with non-trivial 
holonomy, and their monopole constituents as relevant 
degrees of freedom for the confined phase.
\end{abstract}

\maketitle

\section{Introduction}
New instantons (also called calorons) have been obtained recently, 
where the Polyakov loop at spatial infinity (the so-called holonomy) is 
non-trivial \cite{PLB,LeLu}. 
Trivial holonomy, i.e. with values in the center of the gauge group, is 
typical for the deconfined phase \cite{HaSh,GPY}. Non-trivial holonomy is 
therefore expected to play a role in the confined phase (i.e. for $T<T_c$) 
where the trace of the Polyakov loop fluctuates around small values. 

The Polyakov loop plays the role of the Higgs field, $P(t,\vec x)\!=\!\Pexp
\left(\int_0^\beta A_0(t+s,\vec x)ds\right)$, where $\beta=1/kT$ is the period 
in the imaginary time direction. For SU($n$), finite action requires this to 
tend to
\beq
\pl=\lim_{|\vec x|\to\infty} P(0,\vec x)=g^\dagger
\exp(2\pi i\diag(\mu_1,\mu_2,\ldots,\mu_n))g,
\eeq
where $g$ is chosen to bring $\pl$ to its diagonal form, with the $n$ 
eigenvalues being ordered according to $\mbox{$\sum_{i=1}^n\mu_i=0$}$ and 
$\mu_1\leq\mu_2\leq\ldots\leq\mu_n\leq\mu_{n+1}\equiv1+\mu_1$. One can
recognize $8\pi^2\nu_m/\beta$ (with $\nu_m=\mu_{m+1}-\mu_m$) as being the 
monopole mass.

Monopoles as constituents are close to the picture of instanton quarks, which 
was already introduced more than 25 years ago \cite{BFST}. The only 
difference is that instanton quarks were pointlike, whereas here we have 
to work in terms of monopole degrees of freedom. We will investigate
in how far this plays a role in describing confinement.

Caloron solutions are such that the total magnetic charge vanishes. The 
"force" stability of these solutions in terms of its constituent monopoles 
is based, as for exact BPS multi-monopole solutions, on balancing the 
electromagnetic with the scalar (Higgs) force \cite{Mant}, except 
that for calorons repulsive and attractive forces are interchanged as 
compared to multi-monopoles. A single caloron with topological charge one 
contains $n-1$ monopoles with a unit magnetic charge in the $i$-th U(1) 
subgroup, which are compensated by the $n$-th monopole of so-called type 
$(1,1,\ldots,1)$, having a magnetic charge in each of these subgroups. 
At topological charge $k$ there are $kn$ constituents, $k$ monopoles of 
each of the $n$ types. The sum rule $\sum_{j=1}^n\nu_j=1$ guarantees the 
correct action, $8\pi^2 k$, for calorons with topological charge $k$.

\section{One-loop Corrections} 

Prior to their explicit construction, calorons with non-trivial holonomy were 
consi\-der\-ed irrelevant \cite{GPY}, because the one-loop correction gives 
rise to an infinite action barrier. However, the infinity simply arises due 
to the integration over the finite energy density induced by the perturbative 
fluctuations in the background of a non-trivial Polyakov loop~\cite{Weiss}. 
The non-perturbative contribution of calorons (with a given asymptotic value 
of the Polyakov loop) to this energy density as the relevant quantity to be
considered, was first calculated in supersymmetric theories \cite{Khoze}, 
where the perturbative contribution vanishes. It has a minimum where the trace 
of the Polyakov loop vanishes, i.e. at maximal non-trivial holonomy. 
Recently the calculation of the non-perturbative contribution was performed in 
ordinary gauge theory at high temperatures \cite{Diak}. When added to the 
perturbative contribution with its minima at center elements, these minima turn 
unstable for decreasing temperature right around the expected value of $T_c$. 
This lends some support to monopole constituents being the relevant degrees of
freedom which drive the transition from a phase in which the center symmetry is 
broken at high temperatures to one in which the center symmetry is restored at 
low temperatures. 
\begin{figure}
\includegraphics[height=0.17\textheight]{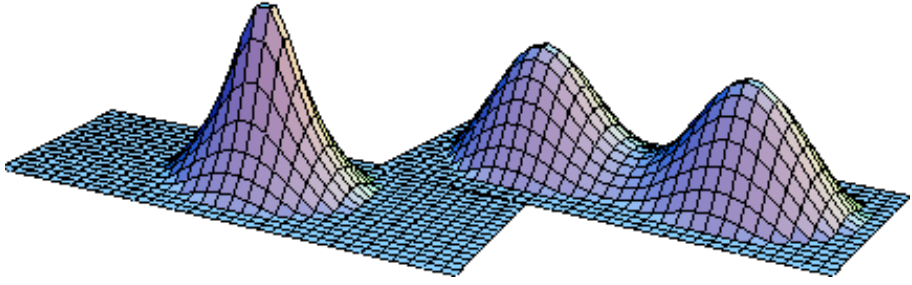}
  \caption{Profiles for a caloron at $\omega=1/4$ with $\rho\ll T$ (left) and 
  $\rho=T$ (right), where vertically the logarithm of the action density is 
  plotted, cutoff below $1/(2e)$.}
\end{figure}

\section{A Caloron Gas Model for Confinement}
 
A caloron gas model has been constructed recently for SU(2) \cite{GIMP}, 
where one solves for overlapping instantons approximately. One takes 
\beq
A_\mu^{\rm per}(x)=e^{-2\pi i\vec\omega\cdot\vec\tau}\sum_i A_\mu^{(i),{\rm 
alg}}(x)e^{2\pi i\vec\omega\cdot\vec\tau}+2\pi\vec\omega\cdot\vec\tau
\delta_{\mu4}
\eeq
to be valid when the density is of the order of 1 fm${}^{-4}$ and size $\rho$
is roughly 0.33~fm. In other words, one adds the caloron gauge fields (with 
the {\em same} $\pl=e^{2\pi i\vec\omega\cdot\vec\tau}$) in the algebraic gauge
$A_\mu^{\rm alg}(x+\beta)=\pl A_\mu^{\rm alg}(x){\cal P}_\infty^{-1}$ in order 
not to change the boundary conditions. Only at the end one transforms to the 
periodic gauge. This has been shown to be exact for multi-calorons \cite{BrvB},
but for the above parameters it is a good approximation for a superposition of
(anti)calorons.

Remarkably this seems to give confinement for $T<T_c$ and deconfinement
for $T>T_c$. In the confining phase one imposes $\omega=|\vec\omega|=1/4$ and
$\Tr\pl=0$, whereas in the deconfining phase one tends to find $\omega=0$ 
(or 1/2) and $\Tr\pl=2$ (one takes into account that $\omega$ only gradually 
becomes 0 or 1/2 with increasing temperature, but we will ignore this here). 
In figure 1 the caloron is shown for $\omega=1/4$, where we contrast 
$\rho\ll T$ and $\rho=T$. Of course $\rho$ is somewhere in between, but 
it clearly gives a confining force over the distances probed.

To show this they have solved for 
\beq
D_1(\rho,T)=A_1\rho^{b-5}\exp(-c\rho^2)\quad{\rm and}\quad
D_2(\rho,T)=A_2\rho^{b-5}\exp(-4[\pi\rho T]^2/3),
\eeq
where in the first case $\bar\rho$ is fixed, $T\leq T_c$ and $\omega=1/4$
(which means $\nu=1/2$), and in the second case $\bar\rho$ is running,
$T\geq T_c$ and $\omega=\nu=0$. Finally one requires $\bar\rho(T_c)_{\rm
conf}=\bar\rho(T_c)_{\rm deconf}=0.37$ fm, which determines $c$. With 
$b=(11n-2n_f)/3=22/3$ ($n_f=0$) and $\int D_{1,2}(\rho,T)d\rho=1$ this 
gives the model. Determining $\bar\rho(T<T_c)$, they have also 
fixed $T_c\approx 178$ MeV and $\sigma(0)\approx318$ MeV/fm.
\begin{figure}
\includegraphics[height=0.3\textheight]{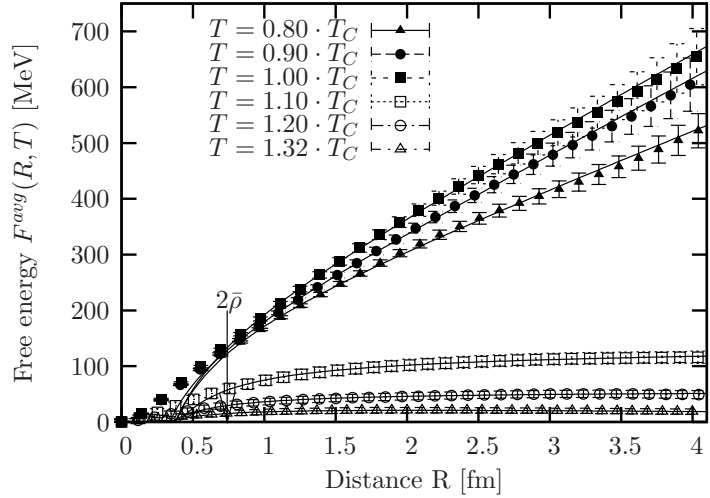}  
\caption{Free energy versus distance $R$ at different 
temperatures ${T/T_C=0.8,\, 0.9,\, 1.0}$ for the 
confined and at ${T/T_C=1.10,\, 1.20,\,1.32}$ for 
the deconfined phase in the fundamental representation.}
\end{figure}

In figure 2 the free energy versus the distances at different temperatures
is given and although the string tension should go to zero as one approaches 
$T_c$ from below, it is true that for $T<T_c$ the string tension is finite and 
becomes zero for $T>T_c$. This model, in a sense, assumes weak coupling.
Also in the spatial Wilson loops one finds an area law.

\section{Dense Matter}

There has been yet another development that introduces instanton quarks
to describe confinement \cite{SSZ}, which has been summarized in \cite{Zhit}.
At low energies and large chemical potential the $\eta'$ interactions are 
determined by ordinary instantons, with a periodicity of $\theta$ which is 
$2\pi$. But at small chemical potential (and temperature) one finds for
$\eta'=\phi=\Tr(U)$, where $U$ is the chiral matrix, that (ignoring the 
mass corrections) 
\beq
L_{\eta'}=f^2(\partial_\mu\phi)^2+\lambda\cos([\phi-\theta]/n).
\eeq
Now the topological charge is $Q_a=\pm 1/n$, but with the sum $Q=\sum_a Q_a$ 
an integer. The conjecture 
is that in the confined phase instanton quarks can be far apart, but remain 
strongly correlated, requiring large and overlapping instantons. One has to 
see if it is strongly interacting and if the constituents are line like  
(the constituent monopoles), instead of point like (at least semi-classically). 
The conclusions are nevertheless interesting.

In conclusion instanton quarks seem to play a role in the confined phase. 
The interpretation is of course different than what was assumed in
\cite{BFST}, where now the time coordinate is replaced in a sense by 
a phase. What remains true is, however, that charge $k$ SU($n$) solutions are 
described by $kn$ lumps of charge $1/n$.

\begin{theacknowledgments}
I thank Manfried Faber, Misha Polikarpov and above all Jeff Greensite
for inviting me to this wonderful place. I thank Michael Ilgenfritz
and Michael M\"uller-Preussker for explaining their work with Phillip
Gerhold, and Falk Bruckmann for taking over when I was away, and from 
which I am determined to come back, if only for the warmly felt wishes 
of many participants. 
\end{theacknowledgments}


\begin{thebibliography}{99}
\bibitem{PLB}T.C. Kraan and P. van Baal, Phys. Lett. B428 (1998) 268 
[hep-th/9802049]; 
Nucl. Phys. B533 (1998) 627 [hep-th/9805168]; 
Phys. Lett. B435 (1998) 389 [hep-th/9806034]. 
\bibitem{LeLu}K. Lee, Phys. Lett B426 (1998) 323 [hep-th/9802012];
K. Lee and C. Lu, Phys. Rev. D58 (1998) 025011 [hep-th/9802108].
\bibitem{HaSh}B.J. Harrington and H.K. Shepard, Phys. Rev. D17 (1978) 2122; 
Phys. Rev. D18 (1978) 2990. 
\bibitem{GPY}D.J. Gross, R.D. Pisarski and L.G. Yaffe, Rev. Mod. Phys.
53 (1981) 43. 
\bibitem{BFST}A.A. Belavin, V.A. Fateev, A.S. Schwarz and Yu.S. Tyupkin, 
Phys. Lett. B83 (1979) 317.
\bibitem{Mant}N.S. Manton, Nucl. Phys. B126 (1977) 525;
C. Montonen and D. Olive, Phys. Lett. 72B (1977) 117.
\bibitem{Weiss}N. Weiss, Phys. Rev. D24 (1981) 475.
\bibitem{Khoze}N.M. Davies, T.J. Hollowood, V.V. Khoze and M.P. Mattis,
Nucl. Phys. B559 (1999) 123 [hep-th/9905015]. 
\bibitem{Diak}D. Diakonov, N. Gromov, V. Petrov and S. Slizovskiy,  
Phys. Rev. D70 (2004) 036003 [hep-th/0404042]; 
D. Diakonov and N. Gromov, Phys. Rev. D72 (2005) 025003 [hep-th/0502132].
\bibitem{GIMP}P. Gerhold, E.-M. Ilgenfrizt and M. M\"uller-Preussker,
An $SU(2)$ KvBLL caloron gas model and confinement, hep-ph/0607315.
\bibitem{BrvB}F. Bruckmann and P. van Baal, Nucl. Phys. B645 (2002) 105 
[hep-th/0209010]; 
F. Bruckmann, D. N\'ogr\'adi and P. van Baal, Nucl. Phys. B666 (2003) 197
[hep-th/0305063]; 
Nucl. Phys. B698 (2004) 233 [hep-th/0404210]. 
\bibitem{SSZ}D.T. Son, M.A. Stephanov and A.R. Zhitnitsky, Phys. Lett. B510 
(2001) 167 [hep-ph/0103099]; 
D. Toublan and A.R. Zhitnitsky, Phys. Rev. D73 (2006) 034009 [hep-ph/0503256].
\bibitem{Zhit}A.R. Zhitnitsky, Confinement-Deconfinement Phase Transition
and Fractional Instanton Quarks in Dense Matter, hep-ph/0601057. 
\end{thebibliography}
\end{document}